\documentclass{PoS}

\usepackage[utf8]{inputenc}

\usepackage{amsmath}
\usepackage{caption}
\usepackage{microtype}

\title{Excited state contamination in nucleon structure calculations}

\ShortTitle{Excited state contamination in nucleon structure calculations}

\author{\speaker{Jeremy Green}, John Negele, and Andrew Pochinsky\\
  Center for Theoretical Physics, Massachusetts Institute of Technology\\
  E-mail: \email{jrgreen@mit.edu}, \email{negele@mit.edu}, \email{avp@mit.edu}}

\author{Stefan Krieg\\
  Bergische Universität Wuppertal\\
  Jülich Supercomputing Centre, Forschungszentrum Jülich\\
  E-mail: \email{s.krieg@fz-juelich.de}}

\author{Sergey Syritsyn\\
  Lawrence Berkeley National Laboratory\\
  E-mail: \email{ssyritsyn@lbl.gov}}

\abstract{Among the sources of systematic error in nucleon structure
  calculations is contamination from unwanted excited states. In order
  to measure this systematic error, we vary the operator insertion
  time and source-sink separation independently. We compute
  observables for three source-sink separations between 0.93 fm and
  1.39 fm using clover-improved Wilson fermions and pion masses as low
  as 150 MeV. We explore the use of a two-state model fit to subtract
  off the contribution from excited states.}

\FullConference{XXIX International Symposium on Lattice Field Theory \\
		 July 10 – 16 2011\\
		 Squaw Valley, Lake Tahoe, California}

\begin{document}

\section{Introduction}

As Lattice QCD calculations of nucleon structure have been performed
with decreasing pion mass, chiral extrapolations have become less able
to reconcile lattice data with experiment \cite{Alexandrou:2010cm,
  Renner:2011rc}. An important source of systematic error is
contamination from excited states
\cite{Capitani:2010sg,Brandt:2011sj}, i.e.\ the failure to isolate the
ground-state nucleon.

In order to compute nucleon matrix elements and form-factors, we
compute nucleon two-point functions $C_\text{2pt}(T,\vec p')$ and
three-point functions $C_\text{3pt}^\mathcal{O}(T,\tau,\vec p,\vec
p')$ on the lattice \cite{Bratt:2010jn}, where $T$ is the Euclidean
time separation between the source and the sink, $\tau$ is the
Euclidean time separation between the source and the operator
$\mathcal{O}$, $\vec p'$ is the sink momentum, and $\vec p$ is the
source momentum. If the nucleon interpolating operator with momentum
$\vec p$ creates $N$ states with energies $E_n(\vec p)$, then the
two-point and three-point functions have the form
\begin{equation}\label{eq:c2c3}
\begin{split}
  C_\text{2pt}(T,\vec p) &= \sum_{n=0}^{N-1} a_n(\vec p) e^{-E_n(\vec p)T} \\
  C_\text{3pt}^\mathcal{O}(T,\tau,\vec p,\vec p') &= \sum_{n,n'=0}^{N-1} \sqrt{a_n(\vec p)a_{n'}(\vec p')}\sum_i M_i^\mathcal{O}F_i^{n\to n'}(t)e^{-E_n(\vec p)\tau-E_{n'}(\vec p')(T-\tau)},
\end{split}
\end{equation}
where $t=-(p'-p)^2$, $F_i^{n\to n'}(t)$ are transition form-factors
for the transition from state $n$ to state $n'$ via the operator
$\mathcal{O}$, and $M$ is determined by kinematics and by the
definition of the form-factors.

\section{Computing form-factors}

The traditional approach for computing nucleon matrix elements and
form-factors is to use a ratio of three-point and two-point functions:
\[ R^\mathcal{O}(T,\tau,\vec p,\vec p') =
\frac{C_\text{3pt}^\mathcal{O}(T,\tau,\vec p,\vec
  p')}{\sqrt{C_\text{2pt}(T,\vec p)C_\text{2pt}(T,\vec
    p')}}\sqrt{\frac{C_\text{2pt}(T-\tau,\vec p)C_\text{2pt}(\tau,\vec
    p')}{C_\text{2pt}(T-\tau,\vec p')C_\text{2pt}(\tau,\vec p)}}, \]
such that in the limit of large $\tau$ and $T-\tau$, where excited
states have negligible contribution, the amplitudes $a_0(\vec p)$,
$a_0(\vec p')$ and the exponential dependences on $\tau$ and $T-\tau$
are canceled. In practice, we compute these ratios for one or more
fixed values of $T$ and $\tau\in[0,T]$, then for each $(T,\tau)$ we
compute the form-factors $F_i(t)$ from an overdetermined fit to the
ratios $R$. Then we produce ``plateau plots'' of $F_i(t)$ versus
$\tau$, for fixed $T$, and look for a flat region where it is hoped
that excited-state contamination is small.

The new approach introduced here is to do a combined fit to
$C_\text{2pt}$ and $C_\text{3pt}$ using an $N$-state model derived
from Eq.~\ref{eq:c2c3}, for a small value of $N$. In order to further
simplify the fit, we assume the dispersion relation $E_n^2(\vec p) =
m_n^2 + \vec p^2$, and define
\[ \tilde F_i^{n\to n'}(t) = \sqrt{a_n(\vec p)a_{n'}(\vec
  p')}F_i^{n\to n'}(t), \] so that for fixed masses, the fit model for
$C_\text{2pt}$ and $C_\text{3pt}$ depends linearly on the fit
parameters $a_n(\vec p)$ and $\tilde F_i^{n\to n'}(t)$.  This enables
the use of an iterative fitting procedure for the masses, where at
each step the larger set of linear fit parameters is solved for
exactly.  Denoting the two-point and three-point functions to which we
are fitting by $C_\alpha$ and the linear fit parameters by $b_i$, we
minimize
\[ \chi^2 = (A_{\alpha i}(\{m_n\})b_i -
C_\alpha)(S^*)_{\alpha\beta}^{-1}(A_{\beta j}(\{m_n\})b_j -
C_\beta), \] where $S^*_{\alpha\beta}$ is our estimate of the
covariance matrix of $\{C_\alpha\}$. Because typically the number of
independent samples that we have is of the same order as the number of
variables $C_\alpha$, we use a shrinkage estimator of the covariance
matrix \cite{Ledoit2004365}, $S^*= (1-\lambda) S + \lambda T$, where
$S$ is the sample covariance matrix, $T$ is its diagonal part, and
$\lambda$ is estimated from the data such that the expected error of
the correlation matrix $R_{\alpha\beta}\equiv
S^*_{\alpha\beta}/\sqrt{S^*_{\alpha\alpha}S^*_{\beta\beta}}$ is
asymptotically minimized \cite{Schafer:2005}.

If the number of states included in the model is less than the number
of states that have non-negligible contributions to $C_\text{2pt}$ and
$C_\text{3pt}$, then each model state will have to account for the
contributions from more than one state in the lattice data. The
best-fit masses in the model will be weighted averages of the masses
of states that contribute to the data. The relevant weights are
different for different correlators \cite{Tiburzi:2009zp}, so when
doing a 2-state fit, we use a single ground-state mass $m_0$ but allow
the excited-state mass in the model to vary, using one mass
$m_1^{(2)}$ for $C_\text{2pt}$ and a different mass $m_1^{(3)}$ for
$C_\text{3pt}$. The difference between these two masses is an
indicator of the importance of omitted states.

\section{Lattice measurements}

We use 2+1 flavors of tree-level clover-improved Wilson fermions
coupled to double-HEX-smeared gauge links \cite{Durr:2010aw}. Results
presented here are from ensembles with lattice spacing $a=0.116$~fm
and pion masses approximately 150, 200, and 250~MeV. At the lightest
pion mass, the lattice volume is $48^4$, and at the other two it is
$32^3\times 48$. In order to study excited-state contamination, we
compute nucleon three-point functions with three different source-sink
separations $T/a=8$, 10, and 12. We focus on isovector quantities to
avoid contributions from disconnected diagrams. We use the standard
nucleon operator, $N_\alpha =
\epsilon^{abc}(u^T_aC\gamma_5d_b)u_{c\alpha}$, where the quark fields
have smearing tuned to minimize excited states seen in the two-point
function, and we use the spin-parity projection
$\Gamma_\text{pol}=\tfrac{1+\gamma_4}{2}\tfrac{1-i\gamma_3\gamma_5}{2}$. All
of the errors below are statistical and are computed using the
jackknife method.

\begin{figure}
\begin{minipage}[t]{0.48\linewidth}
  \centering
  \includegraphics[width=\textwidth]{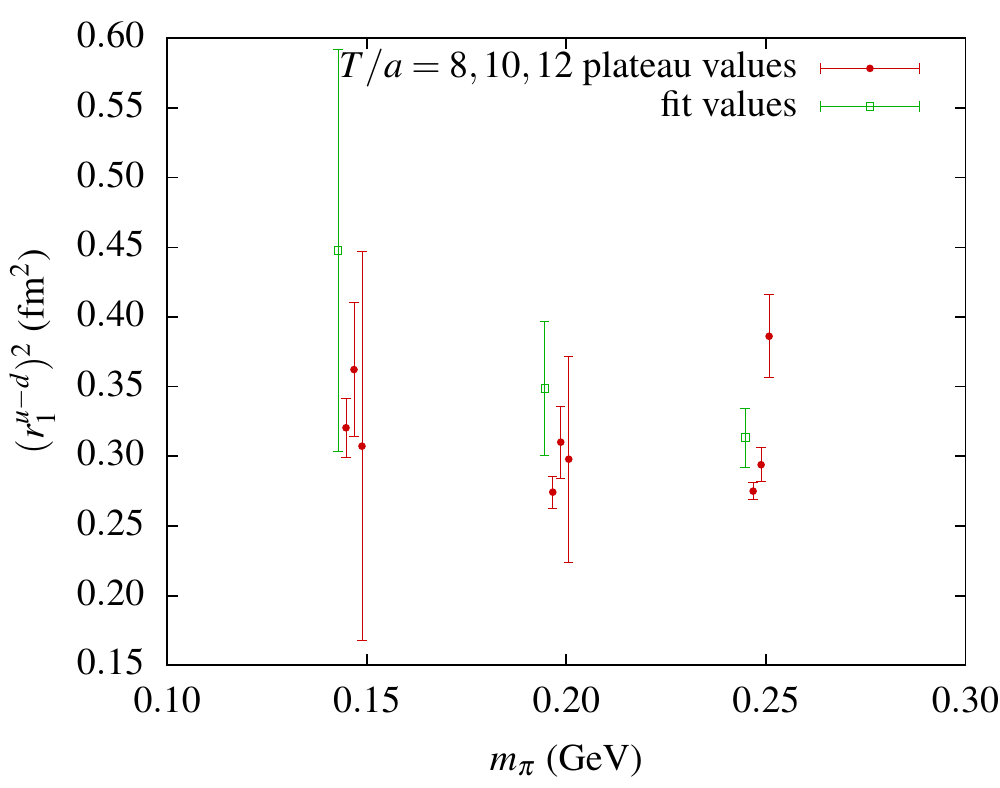}
  \caption{Isovector Dirac radius $(r_1^{u-d})^2$ ($\text{fm}^2$)
    versus $m_\pi$ (GeV). At each pion mass, from left to right are
    results obtained from the fitting procedure, and from the ratio
    method with three increasing source-sink separations.}
  \label{fig:r1sqvsmpi}
\end{minipage}
\hspace{0.03\linewidth}
\begin{minipage}[t]{0.48\linewidth}
  \centering
  \includegraphics[width=\textwidth]{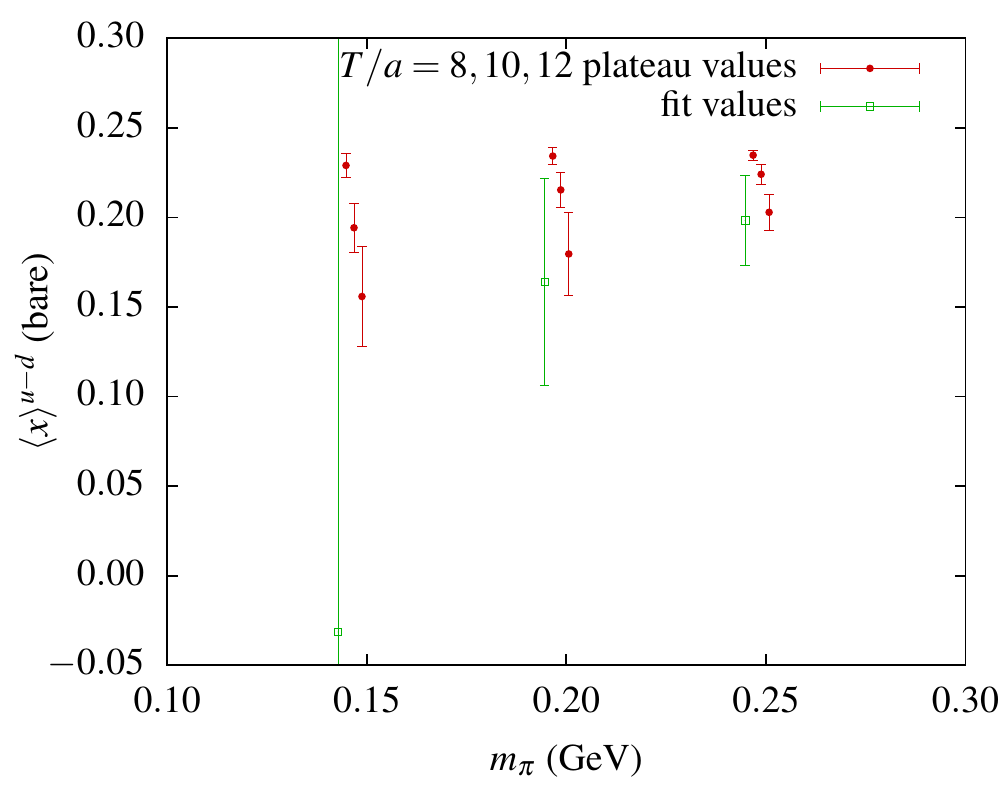}
  \caption{Isovector average momentum fraction $\langle
    x\rangle^{u-d}$ (bare) versus $m_\pi$ (GeV). At each pion mass,
    from left to right are results obtained from the fitting
    procedure, and from the ratio method with three increasing
    source-sink separations.}
  \label{fig:xvsmpi}
\end{minipage}
\end{figure}

Matrix elements of the vector current $V^\mu_q=\bar q\gamma^\mu q$ are
parametrized by the vector form-factors $F_{1,2}(t)$. The mean squared
Dirac radius is defined from the slope of $F_1(t)$ at zero $t$:
$F_1(t) = F_1(0)[1-\tfrac{1}{6}(r_1)^2t + \mathcal{O}(t^2)]$. We
compute the isovector Dirac radius from a linear fit to $F_1(0)$ and
$F_1(t_1)$, where $t_1$ corresponds to three-point functions with
$\vec p$ equivalent to $\tfrac{2\pi}{L}(1,0,0)$, and $\vec
p'=0$. Using the ratio method, we average the three central points of
each plateau plot to arrive at a value of $(r_1^{u-d})^2$ for each
source-sink separation and each ensemble. The results are shown in
Fig.~\ref{fig:r1sqvsmpi}. There is a consistent trend across the three
ensembles of the Dirac radius increasing between $T/a=8$ and
$T/a=10$. However, the behavior going to $T/a=12$ is ambiguous and it
is unclear, from this analysis and with currently available
statistics, whether or not excited-state contamination is still a
problem at $T/a=10$.

We also compute the isovector average momentum fraction $\langle
x\rangle^{u-d}$ from forward matrix elements of the operator
$\mathcal{O}_q^{\mu\nu}=\bar q\gamma^{\{\mu}iD^{\nu\}}q$, where the
braces denote taking the symmetric traceless part of the tensor. Note
that the results presented here have not been renormalized. Again
averaging the three central points of each plateau plot, the results
are seen in Fig.~\ref{fig:xvsmpi}. In this case, a clearer trend is
visible: $\langle x\rangle^{u-d}$ decreases as $T/a$ increases, and
there is no evidence that even $T/a=12$ might be large enough that
excited-state contamination is negligible. This result is consistent
with a recent study using the open sink method, where $\tau$ is fixed
in order to allow the calculation of three-point functions at all
values of $T$ \cite{Dinter:2011sg}. Furthermore, the effect of excited
states appears to be larger at smaller pion masses. This is consistent
with the fact that earlier calculations at larger pion masses using
domain wall valence fermions on an asqtad sea yielded chiral
extrapolations to the physical point in excellent agreement with
experiment \cite{Edwards:2006qx,Hagler:2007xi}.

\section{Fit results}

\begin{figure}
\begin{minipage}[b]{0.48\linewidth}
  \centering
  \begin{tabular}{ll}
    $\chi^2/\text{dof}$ & 89(22)/188\\
    $\lambda$ & 0.036(12)\\
    $m_0a$ & 0.637(6)\\
    $m_1^{(2)}a$ & 1.55(14)\\
    $m_1^{(3)}a$ & 0.99(5)\\
    $a_n(0,0,0)$ & $\begin{pmatrix}2.41(9) \\ 2.87(47)\end{pmatrix}\times 10^{-10}$ \\
    $a_n(\tfrac{2\pi}{L},0,0)$ & $\begin{pmatrix}2.08(8) \\ 2.49(41)\end{pmatrix}\times 10^{-10}$ \\
    $\tilde F_1^{n\to n'}(t_1)$ & $\begin{pmatrix}2.09(9) & -0.18(5) \\ -0.21(4) & -0.23(27)\end{pmatrix}\times 10^{-10}$ \\
    $\tilde F_2^{n\to n'}(t_1)$ & $\begin{pmatrix}7.3(5) & -0.7(2) \\ -0.5(3) & -11.1(7.4) \end{pmatrix}\times 10^{-10}$ \\
%    $F_1(0)$ & 1.094(6)\\
    $F_1(t_1)$ & 0.934(10)\\
    $F_2(t_1)$ & 3.25(13)\\
%    $(r_1^{u-d})^2/a^2$ & 23.3(1.6)
  \end{tabular}
  \captionof{table}{Selected results from fit used to compute
    $(r_1^{u-d})^2$ for the $m_\pi=250$~MeV ensemble. Form-factors
    $F_{1,2}(t_1)$ are not renormalized and are computed from $\tilde
    F_{1,2}^{0\to 0}(t_1)\!\bigg/\!\!\sqrt{a_0(0,0,0)a_0(\tfrac{2\pi}{L},0,0)}$.}
  \label{tab:fit}
\end{minipage}
\hspace{0.03\linewidth}
\begin{minipage}[b]{0.48\linewidth}
  \centering
  \includegraphics[width=\textwidth]{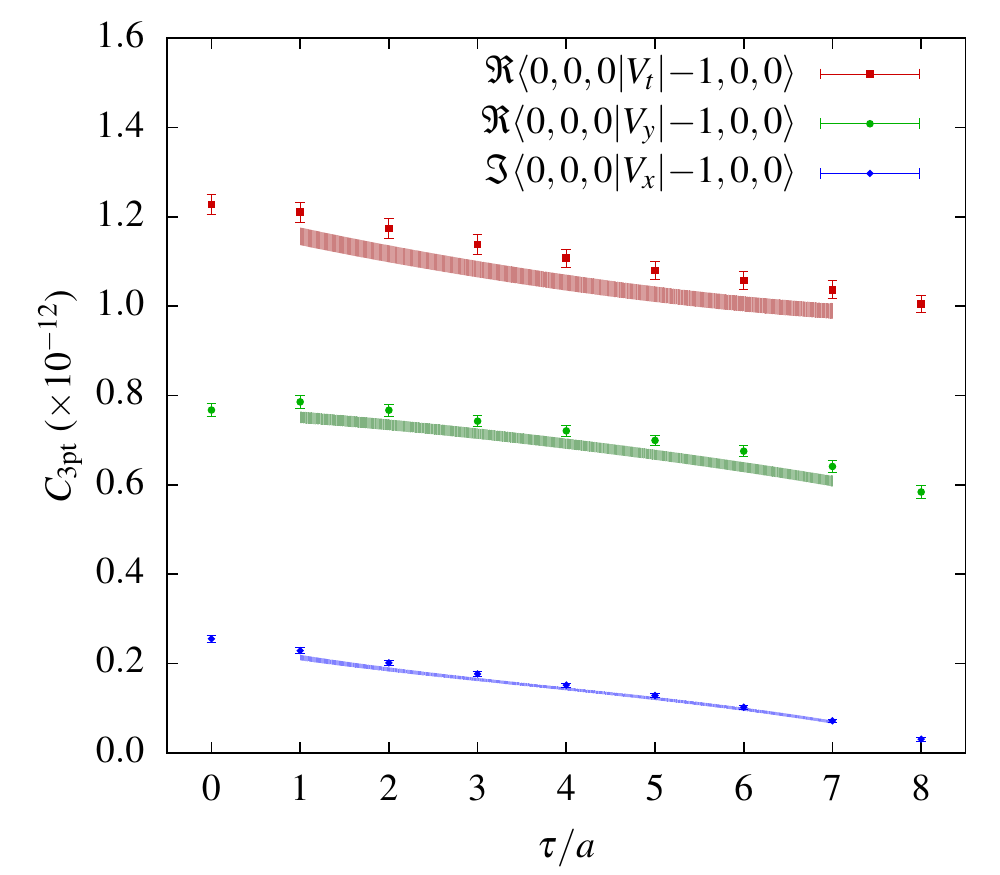}
  \caption{Three-point function (points) and fit (error bands) versus
    $\tau/a$, with $T/a=8$. Matrix element labels are representatives
    from the sets of equivalent three-point functions that are
    averaged to compute the points shown here. Fit bands are
    determined by the quantities $m_0a$, $m_1^{(3)}a$, $\tilde
    F_1(t_1)$, and $\tilde F_2(t_2)$ listed in Tab.~1. Note that the
    points have correlated errors, and that neglecting correlations
    will cause the fit to overlap with the data.}
  \label{fig:fitplot}
\end{minipage}
\end{figure}

We perform 2-state fits to compute $(r_1^{u-d})^2$. This involves fitting to
$C_\text{2pt}$ and $C_\text{3pt}^{V^\mu_{u-d}}$, and the relevant
(transition) form-factors are defined by \cite{Lin:2011da}:
\[ \langle N_{n'}(p',\lambda')|V^\mu|N_n(p,\lambda)\rangle = \bar
u_{n'}(p',\lambda') \left[\left(\delta^\mu_\nu -
    \frac{\Delta^\mu\Delta_\nu}{\Delta^2}\right)\gamma^\nu F_1^{n\to
    n'}(t) + \frac{i\sigma^{\mu\alpha}\Delta_\alpha}{m_n+m_{n'}}
  F_2^{n\to n'}(t)\right] u_n(p,\lambda),\] where
$\Delta=p'-p$. Combining 3-point functions that (up to an overall
sign) have the same set of contributions from form-factors, at the
first nonzero momentum transfer there are three independent
$C_\text{3pt}^{V^\mu}(T,\tau)$, which correspond to the three matrix
elements listed in Fig.~\ref{fig:fitplot}. Including these in a fit
with the three-point functions at zero momentum transfer (both at rest
and boosted), for $\tau/a\in[1,T/a-1]$, $T/a\in\{8,10,12\}$, and with
$C_\text{2pt}(T,\vec p)$, for $\vec p\in\{0,\tfrac{2\pi}{L}(1,0,0)\}$
and $T/a\in[2,12]$, we arrive at a fit to 211 variables, with 20
linear fit parameters and 3 mass parameters $m_0$, $m_1^{(2)}$, and
$m_1^{(3)}$. As before, the Dirac radius is determined using a linear
fit to the resulting $F_1(0)$ and $F_1(t_1)$.

Selected parameters and derived quantities from the fit to the
$m_\pi=250$~MeV ensemble are shown in Tab.~\ref{tab:fit}. In
particular, note that the best-fit excited-state mass for the
three-point function is lower than that for the two-point
function. This suggests that there may be a state that has important
contributions to three-point functions but is not easily detected from
the two-point function of a single smeared nucleon operator. The fit
model is compared with the 3-point functions at the first nonzero
momentum transfer and $T/a=8$ in Fig.~\ref{fig:fitplot}. In this
figure, contributions from the ground-state nucleon will decay
approximately as $\exp(-0.03\tau/a)$, since the nucleon with nonzero
momentum at the source has slightly higher energy than the nucleon at
rest at the sink. This is the approximate behavior seen in the first
two of the three 3-point functions, however the third approaches zero
more rapidly. This can be explained if the different three-point
functions have different relative contributions from excited
states. This is extra information that the fit is able to make use of,
but is discarded when producing a plateau plot.

In Fig.~\ref{fig:r1sqvsmpi}, we compare $(r_1^{u-d})^2$ from this fit
with the values from the ratio method. The fit results show a stronger
trend of increasing Dirac radius at smaller pion masses, although
since the fit points have large errors (only slightly smaller than for
the $T/a=12$ plateau values), the outcome from the fit is consistent
with with the $T/a=10$ plateau values across all ensembles.

To determine $\langle x\rangle^{u-d}$, we perform the analogous fit to
$C_\text{2pt}$ and $C_\text{3pt}^{\mathcal{O}^{\mu\nu}_{u-d}}$, except
that we restrict to $\tau/a\in[2,T/a-2]$, since the operator extends
in the time direction. The (transition) matrix elements are
parametrized by five generalized form-factors:
\begin{align*}
    \langle N_{n'}(p',\lambda')| \mathcal{O}^{\{\mu\nu\}}
  |N_n(p,\lambda)\rangle &= \bar u_{n'}(p',\lambda') \biggl[\bar
    p^{\{\mu}\gamma^{\nu\}}A_{20}^{n\to n'}(t) +
    \Delta^{\{\mu}\gamma^{\nu\}}A_{21}^{n\to n'}(t)\\ &\qquad\qquad + \bar
    p^{\{\mu}\frac{i\sigma^{\nu\}\alpha}\Delta_\alpha}{m_n+m_{n'}}B_{20}^{n\to
      n'}(t) +
    \Delta^{\{\mu}\frac{i\sigma^{\nu\}\alpha}\Delta_\alpha}{m_n+m_{n'}}B_{21}^{n\to
      n'}(t)\\ &\qquad\qquad + \frac{2}{m_n+m_{n'}}\Delta^{\{\mu}\Delta^{\nu\}}C_2^{n\to
      n'}(t)\biggr] u_n(p,\lambda),
\end{align*}
where $\bar p=(p'+p)/2$, and $A_{21}$ and $B_{21}$ vanish for
non-transition matrix elements due to time-reversal symmetry
\cite{Diehl:2003ny}. Including the first nonzero momentum transfer
helps to give a better handle on the excited-state mass. Overall, the
fit has 358 variables, with 38 linear fit parameters and the three
masses. The fit results for $\langle x\rangle^{u-d}\equiv A_{20}(0)$
are compared with the ratio method in Fig.~\ref{fig:xvsmpi}. These
show the same downward trend as the pion mass decreases that was seen
in the $T/a=10,12$ plateau values, but with larger errors. At
$m_\pi=150$~MeV, statistics were too poor, such that the value
obtained from the fit, $\langle x\rangle^{u-d}= -0.03(67)$, is not
useful.

\section{Conclusion}

There is strong evidence of excited-state contamination in nucleon
structure calculations. We see that the effect depends on the
observable and consequently there is a clear need to quantify the
systematic error from excited states for each observable. In
particular, as demonstrated in Fig.~\ref{fig:xvsmpi}, the use of a
single source-sink separation smaller than about 1.4~fm is inadequate
for accurate calculations of the isovector average momentum fraction
near the physical pion mass.

We have presented one promising approach to reducing this problem by
including an additional finite number of excited states in a
fit. Applying this method with one excited state, we obtain results
that, with currently available statistics, are consistent with the
ratio method.

\acknowledgments

We are pleased to acknowledge use of dynamical gauge field
configurations provided by the BMW collaboration. This research was
supported in part by funds provided by the U.S.\ Department of Energy
(DOE) under cooperative research agreement DE-FG02-94ER40818 and
Contract No.\ DE-AC02-05CH11231. Computer resources were provided by
the DOE through the ASCR Leadership Computing Challenge program at
Argonne National Laboratory and through its support of the MIT Blue
Gene/L.

\bibliographystyle{JHEP-2}
\bibliography{excitedstates.bib}

\end{document}